The imaginary and real velocity of an orbiting body based on different types of conics sections

Andika Arisetyawan

Universitas Pendidikan Indonesia

Jl DR Setyabudhi No. 229 Bandung, Indonesia

andikaarisetyawan@upi.edu

**Abstract**. In this paper, I introduce general equation of conics sections based on physical problem on the earth surface in [1]. The conics sections here (hyperbola and ellipse) are generated by all maximum points of parabolas. Based on it, I derived them to calculate the velocity and kinetic energy for different types of conics. The main results showed that there is imaginary velocity if type of conics is hyperbola for $r = 0$, but kinetic energy never be imaginary, only be negative value if $r = 0$. Mean while, if types of conics is ellipse, then the velocity is real and kinetic energy always be positive. It is mathematically unique because only imaginary mass based on special relativity that can produce negative energy.

Keywords: polar coordinate, cartesian coordinate, conics equations

## Introduction

The conic sections in physics play a fundamental role, any body under influence of inverse square law force must have a trajectory that is one of conics section (parabola, ellipse, circle or hyperbola). An interesting case is of the forces of attraction or repulsion between electrically charged particle also obey an inverse square law and such particles also have trajectories that are conics [3]. This paper described about the conics that are generated by all maximum points of parabola. Based on [1], I have derived a formulae from a semicircular motion on the earth surface using Newton's second law of motion. It has a unique form because semicircular motion on the earth surface has similar formulae with parabola but the coefficient of the highest order change from time to time. In this paper, by setting different value for $\theta$, we have different trajectory or geometry. Forgetting about unfamiliar symbols and complicated mathematical technique which is only recognized by experts. It only use simple physics and mathematics.

## Polar and Cartesian coordinate system

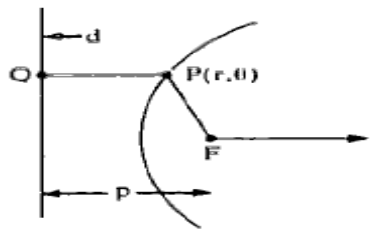

Fig.1 Taken from Kastner (1985)

From the figure above, we rewrite general equation for conics in polar coordinates as follows

$$r = \frac{ep}{1 - e\cos\theta} \tag{1}$$

In which $e$ is eccentricity of conics, $p$ is the distance from the focus to the directrix. Equation (1) can be transformed into cartesian of an ellipse, parabola and hyperbola as follows (for detail, see [3]):

Parabola:

$$y^2 = a(x - b)^2 \tag{2}$$

With:

$$a = 2p, b = -\frac{p}{2}$$

Ellipse :

$$\frac{(x - d)^2}{a^2} + \frac{y^2}{b^2} = 1 \tag{3}$$

With:

$$d = \frac{pe^2}{1 - e^2}, a = \frac{pe}{1 - e^2}, b = \frac{pe}{\sqrt{1 - e^2}}$$

Hyperbola:

$$\frac{(x - d)^2}{a^2} - \frac{y^2}{b^2} = 1 \tag{4}$$

With:

$$d = \frac{pe^2}{1 - e^2}, a = \frac{pe}{e^2 - 1}, b = \frac{pe}{\sqrt{e^2 - 1}}$$

And, we have total energy for conics is

$$E = \frac{GMm(e^2 - 1)}{2ep} \tag{5}$$

In which $e = 0$ and $e = 1$ total energy are infinite and zero. Since it is impossible in real world for total energy to have value that $e = 0$ and $e = 1$, therefore, types of orbits that exactly parabola and circle will not occur in nature (see[3]).

**Semicircular Motion on the Earth Surface**

Based on [1], Arisetyawan has derived semicircular motion from a real problem on the earth surface using Newton's second law of motion as follows (complete derivation formulae can be seen in [1]):

$$y = (\tan\alpha)x - 0.5\left(\frac{v^2 \cos(90 - \theta)}{v_0^2 \cos^2\alpha\, r} + \frac{g(\sin^2(90 - \theta))}{v_0^2 \cos^2\alpha}\right)x^2 \tag{6}$$

We will prove that equation (6) are parabola, ellipse and also hyperbola

If $\theta = 0$, then equation (6) will be

$$y = (\tan \alpha)x - 0.5 \left(\frac{g}{v_0^2 \cos^2 \alpha}\right) x^2 \tag{7}$$

Which is parabola in physics with assumption $g, v_0, \alpha$ are constant.

If $\theta = 90°$, then equation (6) will be

$$y = (\tan \alpha)x - 0.5 \left(\frac{v^2}{v_0^2 \cos^2 \alpha \, r}\right) x^2 \tag{8}$$

A glimpse, it is similar with parabolic trajectory, but we will prove that coefficients of $x^2$ in (8) is not constant for all time $t$. From the relation between angular velocity and time (see [1],[2]), we have

$$t = 1, \omega = \frac{\pi}{2} rad/s, \theta = 90°$$

$$t = 2, \omega = \frac{\pi}{4} rad/s, \theta = 90°$$

$$t = 3, \omega = \frac{\pi}{6} rad/s, \theta = 90°$$

essentially, we can use another constant value for $\theta$, such as $30°, 45°$ for all time $t$. But, we need to remove constant of gravity $g$ to simplify equation (6). So that, for $t = T$, we get

$$\omega = \frac{\pi}{2T} rad/s \tag{9}$$

From the relation between linear and angular velocity, we have

$$v = \omega r \tag{10}$$

from (10), Since $\omega$ is not constant value, then $v$ is not constant too. By considering (8) as parabolic equation, we get symmetrical axis and maximum height as follows:

$$x_s = \frac{v_0^2 \sin 2\alpha}{2 \frac{v^2}{r}} = \frac{v_0^2 \sin 2\alpha}{2a_s}$$

$$\sin 2\alpha = \frac{x_s}{v_0^2 / 2a_s} \tag{11}$$

$$y_{max} = \frac{v_0^2 \sin^2 \alpha}{2 \frac{v^2}{r}} = \frac{v_0^2 \, 2\sin^2 \alpha}{4a_s} = \frac{-v_0^2 (\cos 2\alpha - 1)}{4a_s}$$

$$\cos 2\alpha = -\frac{y_{max} - v_0^2/4a_s}{v_0^2/4a_s} \tag{12}$$

Since $v$ is not constant based on (10), then $x_s$ and $y_{max}$ are not single maximum points.

We can express equation (11) into hyperbolic functions as follows:

$$\sin 2\alpha = \frac{\sinh i\, 2\alpha}{i} = \frac{\sinh \beta}{i} \tag{13}$$

In which $i$ is imaginary number defined by $\sqrt{-1}$.

from (13) and (11), we have

$$\sinh \beta = \frac{i x_s}{v_0^2/2a_s} \tag{14}$$

and from (12), we get

$$\cos 2\alpha = \cosh i\, 2\alpha = \cosh \beta \tag{15}$$

By using a property in hyperbolic functions, equation (14) and (15) can be written as follows

$$\cosh^2 \beta - \sinh^2 \beta = 1 \tag{16}$$

$$\frac{\left(y_{max} - v_0^2/4a_s\right)^2}{\left(v_0^2/4a_s\right)^2} - \frac{(ix_s)^2}{\left(v_0^2/2a_s\right)^2} = 1 \tag{17}$$

$$\frac{(y_{max} - d)^2}{(d)^2} - \frac{(x_s)^2}{(b/i)^2} = 1 \tag{18}$$

$$\frac{(y_{max} - d)^2}{(d)^2} - \frac{(x_s)^2}{(f)^2} = 1 \tag{19}$$

Which is a hyperbola that is generated by all maximum points of parabola.

Equation (18) also can be written into ellipse equation as follows:

$$\frac{(y_{max} - d)^2}{(d)^2} + \frac{(x_s)^2}{(b)^2} = 1 \tag{20}$$

We recall that for a hyperbola in equation (19), $d^2 - f^2 = d^2 + b^2 = c^2$, where $c$ is the distance between the center of the conics and a focus, and $b$ is the length of semimajor axis. Since eccentricity is $c$ divided by $b$, then we have $e = 0.5\sqrt{5}$. it means the eccentricity of hyperbola is constant and it does not depend on $v_0, v$ and $r$. By using similar process for an ellipse, we get eccentricity for an ellipse is $0.5\sqrt{3}$.

from [3], it is known that total energy of a two-body gravitational system related by

$$E = 0.5\,mv^2 - \frac{GMm}{r} = \frac{GMm(e^2-1)}{2ep} \tag{24}$$

Then, for $e = 0.5\sqrt{5}$ (hyperbola),

$$E = 0.5\,mv^2 - \frac{GMm}{r} = \frac{GMm}{4\sqrt{5}p} \tag{25}$$

$$0.5\,mv^2 = \frac{GMm}{4\sqrt{5}p} + \frac{GMm}{r} \tag{26}$$

$$0.5\,v^2 = \frac{GM}{4\sqrt{5}p} + \frac{GM}{r} \tag{27}$$

$$v^2 = GM\left(\frac{1}{2\sqrt{5}p} + \frac{2}{r}\right) \tag{28}$$

$$v = \sqrt{GM\left(\frac{1}{2\sqrt{5}p} + \frac{2}{r}\right)} \tag{29}$$

Calculating focal parameter $p$ from (18),

$$p = \frac{d^2}{\sqrt{d^2+b^2}} = \frac{v_0^2 r}{4\sqrt{5}v^2} \tag{30}$$

Substituting (30) into (29), we get

$$v = \sqrt{GM\left(\frac{4\sqrt{5}v^2}{2\sqrt{5}v_0^2 r} + \frac{2}{r}\right)} \tag{31}$$

$$v^2 = GM\left(\frac{2v^2}{v_0^2 r} + \frac{2}{r}\right)$$

$$v^2\left(1 - GM\frac{2}{v_0^2 r}\right) = \frac{2GM}{r}$$

$$v^2 = \frac{2GM}{r\left(1 - GM\frac{2}{v_0^2 r}\right)} = \frac{2GM}{r - GM\frac{2}{v_0^2}} \tag{32}$$

$$v = \sqrt{\frac{2GM}{r - GM\frac{2}{v_0^2}}} \tag{33}$$

If denominator of equation (33) $< 0$, then $v$ will be imaginary velocity only if $r = 0$
From (32), we can derive kinetic energy as follows:

$$EK = \frac{GMm}{r - GM\frac{2}{v_0^2}} \tag{34}$$

By using similar process for $e = 0.5\sqrt{3}$ (ellipse), we get

$$v = \sqrt{\frac{2GM}{r + GM\frac{2}{v_0^2}}} \tag{35}$$

And kinetic energy will be

$$EK = \frac{GMm}{r + GM\frac{2}{v_0^2}} \tag{36}$$

## Results and discussion

From the results above, we can summarize the differences between hyperbola orbit and elliptical orbit in this paper. In hyperbola orbit, $v$ will be imaginary velocity only if $r = 0$, but for elliptical orbit, $v$ never be imaginary velocity even $r = 0$. Meanwhile, kinetic energy in hyperbola could be negative if $r = 0$, but never be imaginary. On the other hand, in elliptical orbit, kinetic energy always be positive.

Since eccentricity of hyperbola and ellipse are constant and do not depend on $r$, it means the sizes of them are differents for every value of $r$ but the forms are same. Although the forms of hyperbola and ellipse are same for every different value of $r$, but total energy aren't. It depends on the value of $M, m$ and $p$. Since $p$ depends on the value of $v_0, v$ and $r$. Then we can say that total energy depends on the value of $r$.